# Learning Programming : An Indian Perspective

Biswajit Saha
Computer Science and Engineering department
Dr.B.C Roy Engineering College
India
frombiswajit@rediffmail.com

Utpal Kumar Ray
Information Technology Department
Jadavpur University
India
utpal_ray@yahoo.com

*Abstract—* **Rapid changes in the field of software engineering have increased the requirement of designing and developing of extremely complex software systems. These tasks are performed by software engineers. Software engineers working in application development have to deeply analyze users' needs and requirements and build software systems that meet the needs of the users. They must possess excellent problem solving skills. They must also have strong programming skills. While teaching introductory programming courses for over a decade in reputed institutions we have experienced that several factors play significant role in developing problem solving skills and program development skills in students. There are certain types of difficulties that are encountered by the beginners. These difficulties vary in their nature. Beginners find difficulties with the programming language that they use, the compilers that they use and so on. These difficulties if not overcome proves detrimental to their career as professional engineers at a later stage. This paper focuses on the various types of difficulties that a novice programmer faces while learning programming and tries to find out ways to overcome them.**

## I. INTRODUCTION

The process of learning computer science as a discipline is far removed cognitively from other disciplines because it involves a level of thought that has no real world counterpart. Computers are man made machines that harness and process information. They are designed to think for us, yet we are the ones who program them to think. It is the manner in which information is represented in computers that causes a great challenge to its understanding. Engineering curriculums provide courses for students that support introductory programming. Novices or fresh students who are new to programming struggle initially in their attempts to grasp various concepts involved in programming. Programming languages taught in introductory programming courses abstract the underlying hardware and operating system by providing a high-level syntax, analogous to operations encountered in the real world, to control computer resources. However the extent of what can be considered analogous is usually limited to mathematical and relational operations. Concepts such as declaration of variables, the association of data types with variables, the initialization of variables, looping constructs, functions and recursion are things that people do not take into account when going about making decisions in their daily routines. These concepts are intuitive to the thought process, but need to be explicitly represented in a program. This is where the difficulties encountered by novice programmers begin. They assume too much or too little about the reasoning ability of a computer and most of the times unsuccessfully attempt to rationalize with it. Novice programmers face difficulties in breaking down a given problem, designing a working solution, writing and debugging a program. Research in the area of program development has been conducted as long as computers have existed. Research in the area of problem solving has been going on for centuries, but only in the past fifty years has it been considered in the context of programming. The results of research have provided valuable input in improving the pedagogy of problem solving and program development, methodologies, paradigms, programming language design and the development of novice learning tools.

## II. DIFFICULTIES FACED BY NOVICES

The following barriers have been identified as hindrances to learning: lack of motivation; limited knowledge in science; inability to understand abstract relationships and difficulty keeping track of a complex discussion.

### A. Pedagogical Roots

Mode of teaching in an introductory course is one of the most vital factors that influence how a novice programmer will perform in higher-level courses. If instructors fail to teach the required skills or if the novices fail to acquire the desired skills many first year students may get discouraged as problems get harder to solve. It may eventually lead them away from the programming aspect of computer science and software engineering.

*1) Problem Solving:* A major problem area attributed to pedagogy is the teaching and learning of problem solving skills. This area is generally neglected in the first year programming course. When first year students are assigned to write programs, they frequently generate the source code without any organized thought process. They proceed iteratively by modifying their source code as required until the program generates the correct outputs. Students begin to type their programming solutions seconds after the initial reading of a problem statement. This habit generally stays even when they study in the higher semester.

*II) Program Comprehension:* Program comprehension is the process that facilitates the understanding of an existing program. This may also lead to the program being reused for a higher-level task. Program comprehension is very difficult for novices. They read the program one line at a time. They view a computer program similar to a cookbook recipe with the expectation that all program statements are executed one after another, in exactly the same way as they appear in the program.

*B. Psychological Roots*

In the following sub-sections some of the difficulties are discussed.

*I) Design of Programming Languages:* The psychologies of problem solving and program development have been very widely and extensively research areas in computer science. However, studies and findings in these areas seem to have very little effect on the design of programming languages. The development of programming languages focuses primarily on advances in technology and practically discards psychological dimensions.

*II) Human Computer Interaction:* According to reference [1] designers of programming languages are highly intelligent experts in the field of programming and are consequently far removed both cognitively and temporally from the difficulties experienced by novice programmer. As a result novices have to adjust their thinking skills in ways that are unnatural to them. The variance between how programmers perceive a solution and the way it must be syntactically expressed in a programming language affect beginners who are just learning to program. Novice programmers find it difficult to conceptualize what is meant by a variable assignment, how they are stored in memory and the lifetime of a variable. Some view a variable as an imaginary box that stores any type and amount of data and even expect the computer to figure out the intent of a variable based on its name. They have difficulty understanding the concept of memory organization within a computer. Due to which, frequent problems like violations of boundary and range conditions, precision errors and calculations that result in computational overflow and underflow occur. Students are surprised to find that the addition of two positive integers may sometime result in a negative value. Another problem that novices face is variable initialization. Some take time to understand why a variable should be initialized prior to its use. Most of these misconceptions are based on mathematical experiences. In solving mathematical problems on paper, one does not have to deal with constraints imposed by a computer and subsequently will not have to deal with the issues mentioned above. Lack of proper program development environments play a significant role in originating misconceptions. There are very few environments that animate data flow and control flow during the execution of a program in real time. Without the use of such a visual tool, novice programmers have a difficult time visualizing the execution of compiled source code residing in the main memory of a computer. Novice programmers also feel that it would have been much better if the compiler or the computer that they use would have understood natural language and associated syntax. That would have made programming much easier for students new to programming. They also expect a computer to interpret a single program statement just as they would. But these are not possible.

*III) Knowledge Acquisition:* Reference [2] identifies the human cognitive system as being divided into two types of memory *short term* and *long term*. He hypothesizes three steps for meaningful learning to occur. The first step is **reception**. In this step the learner must pay attention to the incoming information. The second step is **availability**. Here the learner must process the appropriate prerequisite concepts in long term memory to absorb the new information. The last step is **activation**. In this step the learner must use the prerequisite knowledge so that new material can be connected with it. Since novices do not possess the required prerequisite knowledge about problem solving and program development they are forced to memorize. Given this deficiency it is clear that the pace of the programming course can dictate how effectively a student learns. A lot depends on the instructor of the programming course. This can quickly become a problem if not solved timely because the student will continue to lag behind. This deficiency of knowledge will harm him in many ways.

*IV) Impact of Heuristics on Cognition:* Reference [3] discusses human – computer interaction principles or heuristics that apply to programming such as consistency, simplicity, speaking the user's language and error prevention. More specific guidelines such as closeness of mapping, viscosity, hidden dependencies, imposed guess-ahead and visibility also provides a good foundation for assessing programming systems. But unfortunately it is not easy to optimize all of these aspects at once because of their interdependence. By improving a particular feature there may be reduction in the performance of another feature. As an example, simplifying the syntax of a programming language to that of a natural language may increase its viscosity, the level of effort required to make a small change in the program. Among these principles, reference [4] has identified three principles that seem to cause the most problems encountered in program development: **visibility, closeness of mapping** and **speaking the user's language.**

Visibility is concerned with how much memory a programmer can process and retain at a time. Closeness of mapping determines the level of effort required to transform a mental plan into one that a computer can understand. The programming language syntax greatly affects this. Minimizing this difficulty of transformation may help novice programmers. It is important that the programming languages speak the user's language. When languages use words and symbols, which are unfamiliar to the users, or which have conflicting meanings in other domains, it is hard for novices to learn their purpose [4].

## C. Programming Languages

Nowadays, in India undergraduate engineering curriculums favor the teaching of C or C++ in introductory programming courses. These language and others like them, may not be a good choice for first year students of engineering mainly because of their powerfulness and abundant features. Reference [1] identifies that dichotomies of perspective, such as syntax and semantics, static and dynamic structure, process and data complicate the teaching of any programming language. The availability of low-level implementation oriented constructs and high level solution oriented features in a single language only serves to increase substantially the already considerable cognitive demands placed on the student. One approach to improve the understanding of language syntax and semantics is to design flexible programming languages that appear more natural to the target group of students. However, it is equally important that striving for naturalness does not mean programming language to use natural language. In the following sub sections some difficulties that the target groups of students face while learning to program are discussed.

*I) Issues Involving Syntax:* In a study of novice solving paths conducted by reference [5], a conclusion was drawn that programming language syntax can be a negative factor in problem solving. Any engineering department's choice, especially that of departments like computer science, information technology and software engineering, of a first programming language is very important in building the self confidence level of first year students. The programming language taught must not be too complex. A wide range of features necessitates a commensurately complex syntax and often also entails a host of implicit operations and function calls, automatic conversions, type inferences and resolutions of overloaded functions, variable and function scoping [1]. There are programming languages, which have a simple and primitive syntax. In such languages also syntax may be considered to be cryptic or not understandable to novices as complex syntaxes are. An example of this is the early LISP commands, car and cdr. A syntax that provides operations that closely resemble other operations can pose problems for novices. As for example in the C programming language, a common problem is the use of the assignment (=) and equality (= =) operators. Novices make mistakes whenever they use these operators. They mistakenly interchange these two operators frequently.

*II) Semantics and Pragmatics:* Semantic knowledge refers to understanding the underlying mechanics of syntactical constructs. Pragmatic knowledge refers to understanding why a certain syntactic construct is used in lieu of another even though they may achieve the same end result. Pragmatic knowledge is gained through experience and is something that novices do not have because of their lack of programming experience. At times, they even do not know that an alternative exists. When alternative exists, they often cannot make judgements as to which method is more applicable. As for example, the use of recursion vs. iteration. Since recursion can be very difficult for beginners to understand, they frequently resort to the use of iteration. They may understand that recursion in certain situations may out perform iteration but choose not to use because of its complex nature. Vacillation to use a syntactical construct such as recursion is harmful because some solutions can only be achieved through the use of recursion [5].

## D. Debugging Skills

Debugging a program involves four aspects. At first, a programmer must identify and correct all syntactical errors. Once the program is syntactically correct, the student can begin to test the functionality of the program. During testing he may encounter two types of errors. They are **run time errors** and **logical errors**. Once the sources of these errors are identified, the student makes the appropriate corrections and then re-evaluates the results. Finally, when the program seems to be functioning properly, he may choose to rewrite certain areas of source code in the interest of optimizing performance. Much of the debugging skills are learned through the experience of writing programs. Since novices lack program comprehension skills debugging is not very easy for them. Novices also unintentionally inject errors of their own into the source code while debugging. This happens due to drawing incorrect conclusion about what the error may be or attempting to correct a potential error without examining all possible side effects of doing so. Another frequent cause of this is that students repeatedly attempt to correct their errors without understanding the meaning of the error message produced by the compiler [5].

## E. Compilers

The complexity level of a compiler greatly influences a novice programmer. So, most introductory programming courses' instructors advise their students to use compilers that are relatively simple and easy to use. However, some students who access popular compilers such as Microsoft Visual C++ have a difficult time learning to use it. The focus is then drawn from how to program correctly to how to use a complex compiler. When a compiler finds an error, an indication of what type of error occurred along with the line number of the error producing source code statement is usually presented on the screen. The error messages that they present may be beyond the scope of a novice student. As a result, novice programmers have to deal with full syntax and semantics of the language, even though they do not understand most of it. A compiler can also be a very beneficial tool in a novice's learning experience. Once novices learn how to navigate through the functions of a compiler, they can learn to harness the information provided by the compiler and apply it to existing knowledge [5].

### III. SUGGESTED SOFTWARE ENGINEERING TOPICS IN INTRODUCTORY COURSE

The need for learning some basic software engineering principles early in the course is very necessary.

*A. Importance of Software Engineering*

Early programmers used an exploratory coding style. The exploratory development approach breaks down when the program size increases. A major part of the problem lies in the exponential growth in the perceived complexity and difficulty with program size if one attempts to write the program for a problem without suitably decomposing the problem. As a result, the time, cost and effort required to develop a software product also grow exponentially. In reality, a major emphasis of every software design technique is all about effective decomposition of the large problem into small manageable parts. This also reduces complexity. Handling complexity in a software development problem is a central theme of software engineering discipline. Besides this, there are techniques like problem analysis, software specification, user-interface development, testing and so on [6]. All these collectively help students to be good software engineers in their professional life.

*B. Programs vs Software Products*

Students write programs according to the programming assignments given to them by their instructors. Their programs are small in size and have limited functionality. Also, since the author of a program himself uses and maintains his programs, these usually lack good user-interface and proper documentation. On the other hand, software products have multiple users and have good user-interface. In addition a software product, besides consisting of the source code also consist of all the associated documents such as the requirements specification document, the design document, the test document, the users' manual and so on. A software product is systematically designed, carefully implemented and thoroughly tested. A software product is often too large to be developed by any single programmer. They are developed by group of engineers working in a team and having different jobs to perform. Since a group of software engineer work together in a team to develop a software product, they adopt systematic development strategy. Otherwise, they would find it too difficult to understand each other's work, produce a coherent set of documents and so on [6].

*C. Data Flow Diagram*

Data flow diagrams (also called data flow graphs) show the flow of data through the system. They are very useful in understanding the system under investigation and can be effectively used during system analysis. Beginners should be encouraged to draw DFDs so that their problem solving skills increases.

*D. Data Dictionary*

A data dictionary plays a very important role in any software development process because of the following reasons [6]:
1. A data dictionary provides a standard terminology for all relevant data for use by all engineers working in the same project. A consistent vocabulary for data items is very important, since in large projects different engineers have a tendency to use different items to refer to the same data, which causes confusion.
2. The data dictionary provides the analyst with a means to determine the definition of different data structures in terms of their component elements.

In the data flow diagrams they should be taught to clearly draw the flow of data that takes place in their programs. In the data dictionary they should be taught to write about the objective of the current program, different data types that they use in the program, the different variable names and functions that are used in the program and what purposes they serve. In case of writing programs using functions, they should also mention the actual parameters and formal parameters that they use in the program. If they use any looping construct they should also mention about it and explain its purpose and so on. This may lead to analytical skill development in beginners and they may now be better equipped to face challenges of problem solving.

*E. Structure Charts*

There are two approaches of Software System Design: **Function-oriented design** and **object- oriented design** [7]. In case of function-oriented design, the design can be represented graphically by structure charts. The structure chart of a program is a graphic representation of its structure and the focus is on the hierarchy of the modules.

*I) Modularity*

In case of software design, for solving complex problems the goal is to divide the complex problem into smaller and simpler parts that can be solved separately. These parts or pieces should be able to communicate with each other to solve the large and complex problem. Modularization usually involves the creation and use of small programs which may be referred to as modules.

*II) Flow Chart vs. Structure Chart*

A flow chart differs from a structure chart in three different ways [7]:
1. It is usually difficult to identify different modules of the software from its flow chart representation.
2. Data interchange among different modules is not represented in a flow chart.
3. Sequential ordering of tasks inherent in a flow chart is suppressed in a structure chart.

Beginners should be asked to draw structure charts prior to writing programs so that they get accustomed with the concept of module based program writing. This will also help them to understand better, concepts like **"Component Based Software Engineering"** and **"Software Reuse"** later in their course. So module based program writing should be encouraged among the beginners and programming assignments be given such that they understand the need of **"Software Reusability"** and **"Software Component"**, as early as possible in their course. Moreover this may lead to analytical skill and design skill development much earlier in their career and they may be better equipped to face the challenges of the software industry. It has been observed that beginners have a propensity to underutilize modularization. This problem has its roots partly in the nature of programming assignments that are given to them. Most programming instructors assign programs that have little or no evidence to earlier assignments. Students are not being trained to reuse source code that they have already written because of this strategy of their programming instructors. Instructors need to change this habit to have the full benefit of module based program writing.

*F. Coding Style*

In this section, some concepts related to good coding style in a programming language independent manner are discussed.

*I) Coding Style to Follow*

Good coding style is independent of the programming language in which a program is to be written to solve a problem. The following are some of the guidelines which if followed will help beginners develop good coding style [8]:

- **Names**: Selecting module and variable names is often not considered important by beginners. It is a bad practice though often followed by beginners to choose cryptic and unrelated names. It is also a bad practice to use the same name or a slight modification of it for multiple purposes. Variable names should be closely related to the entity that they represent and module names should reflect their activity.
- **Control Constructs**: As much as possible single entry, single-exit constructs should be used. Also a few standard control constructs should be used rather than using a wide variety of them just because they are available in the programming language.
- **Goto**: Only when the alternative to use goto's is more complex it should be used. Use of goto's in forward transfer is more acceptable than a backward jump. Use of goto's for exiting a loop or for invoking error handlers is also quite acceptable.
- **Information Hiding**: Information hiding though a difficult concept for beginners should be supported wherever possible. Only the access functions for the data structures should be made visible while hiding the data structures behind these functions.
- **User-Defined Types**: Modern programming languages allow users to define new data types like the enumeration type. When such facilities are available, they should be exploited where applicable. Use of this enumerated type makes the program much clearer than defining codes for each day and then working with such codes.
- **Module Size**: A programmer should carefully examine any function (module) with very few lines of source code (say fewer than 7) or with too many lines of source code (say more than 70). Large modules often will not be functionally cohesive and too small modules might incur overhead. There can be however no exact rule specifying the minimum and maximum size of modules. **The guiding principle should be high cohesion and low coupling**.
- **Module Interface**: Any module with a complex interface should be carefully examined. Generally, any module whose interface has more than five parameters should be carefully examined and should be broken into multiple modules with a simpler interface if possible.
- **Program Layout**: Readability of program depends on how the program is organized and presented. Proper indentation, blank spaces and parentheses should be used to enhance the readability of programs. It is a good practice to have a clear layout of programs.
- **Robustness**: A program is robust if it does something planned even for exceptional conditions. A program might face exceptional conditions in such forms as incorrect input, the incorrect value of some variable and overflow. A program should try to handle such situations. Generally, a program should check for validity of inputs where possible and should check for possible overflow of the data structures. If such situations do arise, the program should not just "crash"; it should produce some meaningful message and exit gracefully.

*II) Coding Style To Avoid*

In this sub section some guidelines are given that programmers of all level should follow. These deal with that part of coding style, which should be avoided. They are [7]:

- **Shortcut Methods**: Programs should be written as simply as possible. Cleverly writing programs decreases program readability so much so that even the programmer who wrote cleverly a program might fail to figure out what he has written weeks later. So beginners should be always directed to avoid any shortcut methods while writing programs.
- **Nesting**: Deep nesting should be avoided. Emphasis should be given to construct alternative program segments to avoid deep level of nesting.
- **Side Effects**: When a module is invoked, it sometimes has side effects of modifying the program state beyond the modification of parameters listed in the module interface definition. For example: modifying global variables. Such side effects should be avoided where possible and if a module has side effects, they should be properly documented.
- **Sub optimize**: Sub optimization occurs when one devotes inordinate effort to refine a situation that has little effect on the outcome. The programmer may waste time and effort worrying about situations that are handled in non intuitive ways by the system and that have little effect on overall performance.
- **Identifier**: Programmer sometimes uses one identifier to denote several different entities. The rationale is memory efficiency. Using an identifier for multiple purposes is a bad practice. Since it makes the source code very sensitive to future modifications. Finally, use of an identifier for multiple purposes can be extremely confusing to the reader of a program.

The above are some bad coding styles that beginners should not follow. If they are instructed to avoid the above guideline right from the beginning of their course they will be highly benefited and be able to write source code that is simple and clear even for complex problems.

*III) Coding Standards and Guidelines*

Coding standards are specifications for a preferred coding style. A programming standard might specify items like:
1. Goto statements will be rarely used.
2. Program nesting depths will not exceed 4 levels.
3. Function length will not exceed 40 lines.

A guideline will rephrase these standards in the following manner:
1. The use of goto statements should be avoided in normal circumstances.
2. The nesting depth of programs should be 4 or less in normal circumstances.
3. The number of executable statements in a subprogram should be kept within 40 lines.

So it is very important that students follow certain good coding practices while writing their programs. This will certainly help them in their career in the software industry.

*G. Documentation*

Documentation of software product is an important part of good software development practice and is widely followed in the software industry. This includes all the documents generated during analysis, design, implementation, testing and maintenance of the system. Good documents serve the following purposes [6]:
1. Enhances understandability and maintainability of a software product.
2. Helps the users in exploiting the system.
3. Helps in effectively overcoming the manpower turnover problem.
4. Helps the manager in effectively tracking the progress of the project.

Software documents can be broadly classified into the following:
1. Internal documentation.
2. External documentation.

*I) Internal Documentation*

Internal documentation is provided through appropriate module headers and comments embedded in the source code. It is also provided through the use of meaningful variable names, module and function headers, code indentation, code structuring, etc. Careful experiments suggest that out of all types of internal documentation, the practice of assigning meaningful variable names is the most useful one in understanding the code [6].

*II) External Documentation*

External documentation is provided through supporting documents such as users' manual, software requirements specification document, design document, test documents, etc. A systematic software development style ensures that all these documents are produced in an orderly fashion [6].

Thus it is clear that if students are taught the necessity of documentation early in their courses and instructed to develop the habit of maintaining at least internal documentation while writing their programs they will be benefited.

## IV. CONCLUSION

As mentioned already in the introduction the demand for graduates well trained in software engineering principles and practices is continuing to increase. A late exposure leaves students with very little time to master and practice these principles before they graduate and join the software industry. So it is very important that in today's highly competitive job market, students be taught the right skill timely so that they can do their job in the software industry more confidently, having got enough time to master various skills of software engineering right from the beginning of their undergraduate courses.

Textbooks currently followed for introductory programming courses do not include anything about the topics of software engineering that are suggested to be taught to the beginners. Textbooks written for introductory programming courses should be rewritten to include the suggested topics of software engineering which are some basic principles of software engineering. This will definitely help novice programmers to understand programming better. The textbooks should give solved programming examples and at the same time the data flow diagrams and the structure charts of the solved programming examples should also be shown. The relevant data dictionaries should also be shown.

Technical advances in the field of information technology are very fast. Students belonging to the computer science, information technology and software engineering branches of engineering should be taught latest technology as much as possible. Members of the syllabus committee should closely watch all technical advancements in the field of information technology. Based on the information so obtained they should periodically review the syllabus.

## ACKNOWLEDGMENT

We would like to thank all our students and colleagues without whom this paper would not have been written.